\documentclass[runningheads]{llncs}

\usepackage[T1]{fontenc}
\usepackage{amsmath}
\usepackage{amssymb}
\usepackage{booktabs}
\usepackage{graphicx}
\usepackage{listings}
\usepackage{xcolor}
\usepackage{tikz}
\usepackage{array}
\usepackage{multirow}
\usepackage{url}
\usepackage{microtype}
\usepackage{etoolbox}
\usepackage{mathpartir}
\usepackage{placeins}
\usetikzlibrary{arrows.meta,positioning,shapes.geometric,fit,backgrounds}
\usepackage{cite}
\usepackage{algorithmic}
\usepackage{graphicx}
\usepackage{textcomp}
\usepackage{graphicx}
\usepackage{tcolorbox}
\usepackage{placeins}
\usepackage{listings} 
\usepackage{xcolor}   
\usepackage{hyperref}
\usepackage{everypage}
\usepackage{tikz}

\AddEverypageHook{%
  \begin{tikzpicture}[remember picture,overlay]
    \node[anchor=north,yshift=-0.5cm] at (current page.north) {\small\textit{Accepted in the 19th International Conference on Verification and Evaluation of Computer and Communication Systems (VECoS 2026)}};
  \end{tikzpicture}%
}

\lstdefinelanguage{Coq}{
  morekeywords={Inductive,Type,forall,Theorem,true,false,inr},
  sensitive=true,
  morecomment=[s]{(*}{*)}
}

\newcommand{\encore}{\textit{Encore!}}
\newcommand{\rocq}{Rocq}
\newcommand{\code}[1]{\texttt{#1}}
\newcommand{\derive}[1]{\texttt{\#\allowbreak[derive(#1)]}}

\begin{document}

\title{From Rocq to Metal: A Pipeline for Formally Verified Microcontroller Firmware}
\titlerunning{From Rocq to Metal}

\author{Valentin Bergeron\inst{1} \and Karolina Gorna\inst{2}}
\authorrunning{V. Bergeron and K. Gorna}
\institute{
Ledger, Paris, France\\
\email{bergeron.valentin@gmail.com}
\and
T\'el\'ecom Paris and Ledger Donjon, Paris, France\\
\email{karolina.gorna@[telecom-paris,ledger].fr}
}

\maketitle

\begin{abstract}
Enforcing invariants in safety-critical firmware is increasingly urgent as generated code becomes widespread, but standard extraction targets for proof assistants require runtimes too large for many embedded devices.
We present a pipeline for running formally verified Rocq firmware logic on Cortex-M microcontrollers.
The pipeline extracts Gallina to Scheme, compiles it with \encore, a bare-metal Continuation-Passing Style (CPS) bytecode virtual machine, and embeds the result in \code{no\_std} Rust firmware.
We structure applications as pure state-transition functions, so the business logic is proved in Rocq while the event/effect boundary, host callbacks, compiler, and VM remain explicit trusted components: no proof yet relates bytecode execution to Gallina semantics.
On eight firmware workloads, from APDU parsing to a credential store, \encore{} runs unmodified Rocq-extracted code within the 50\,KiB RAM budget of ST33-class targets on every input size, where CertiRocq's generated code runs out of memory on the larger ones, at the price of up to 10$\times$ more executed instructions than CertiRocq; a transaction-signing application is validated on physical Ledger Flex hardware.

\end{abstract}

\section{Introduction}

Generated code is changing how software is written, including firmware.
For safety-critical devices, lower code-generation cost increases the amount of code that must be reviewed, while the required assurance level remains high.
Formal verification has already demonstrated its value for operating-system kernels, and similar rigor is increasingly sought for firmware.
Formal specifications offer a scalable response: generated implementations can be checked against theorem statements, and proof obligations become part of the development interface.

However, the hardware constraints of embedded firmware prevent straightforward use of standard specification language extraction.
The runtimes required to execute extracted code exceed the resource budget of most microcontrollers.
Standard practice is therefore to translate specifications into the host language, automate that translation, or use host-language extensions, weakening the connection to the proof assistant.

We built \encore{}~\cite{encore} to close this gap.
\encore{} is a bare-metal CPS bytecode VM written in \code{no\_std} Rust that executes Rocq-extracted Scheme on microcontrollers.
The pipeline covers the whole path from proof to device (Fig.~\ref{fig:pipeline}).
We write firmware logic and prove it correct in Rocq, extract it to Scheme, compile it with \encore, and deploy it.
The host Rust application embeds the \encore{} VM as a library, and its \code{build.rs} script runs the \encore{} compiler.
What runs on the MCU is directly derived from the Rocq source, though the compiler, VM, host callbacks, and device I/O remain trusted.

\begin{figure}[t]
\centering
\begin{tikzpicture}[
  font=\scriptsize, >=Latex,
  box/.style={draw, rounded corners=2pt, align=center, inner sep=2pt, minimum height=0.62cm},
  grp/.style={draw, rounded corners=3pt}
]
  \node[box, minimum width=2.4cm] (rocq) at (0.1,0) {\rocq{}\\(Gallina code)};
  \node[box, minimum width=2.0cm] (scm) at (3.4,0) {Scheme code};
  \node[box, minimum width=2.6cm] (cps) at (6.8,0) {Continuation-passing\\style (CPS)};
  \node[box, minimum width=1.6cm] (opt) at (10.1,0) {CPS\\optimizer};
  \draw[->] (rocq) -- node[above, font=\tiny] {extraction} (scm);
  \draw[->] (scm) -- node[above, font=\tiny] {frontend} (cps);
  \draw[->] ([yshift=1.2mm]cps.east) -- ([yshift=1.2mm]opt.west);
  \draw[<-] ([yshift=-1.2mm]cps.east) -- ([yshift=-1.2mm]opt.west);

  \draw[grp] (-1.1,-0.95) rectangle (10.9,-4.05);
  \node at (4.9,-3.82) {\textbf{Microcontroller OS}};
  \draw[grp] (-0.9,-1.15) rectangle (9.0,-3.55);
  \node at (4.05,-3.3) {\textbf{Host} (Rust application)};

  \draw[grp] (-0.7,-1.35) rectangle (3.6,-2.15);
  \node at (1.45,-1.75) {Host application (\code{main.rs})};
  \draw[grp] (-0.7,-2.25) rectangle (3.6,-3.05);
  \node[align=center] at (1.45,-2.65) {\encore{} generated symbols \\(via \code{build.rs})};

  \draw[grp] (4.8,-1.35) rectangle (8.8,-3.05);
  \node[box, minimum width=2.2cm, minimum height=0.55cm] (bc) at (6.8,-1.85) {Bytecode};
  \node[align=center] at (6.8,-2.6) {\textbf{\encore{} VM}\\(library)};

  \draw[grp] (9.2,-1.9) rectangle (10.7,-2.8);
  \node[align=center] at (9.95,-2.35) {Other\\applications};

  \draw[->] (cps) -- node[right, pos=0.27, font=\tiny] {\encore{} compiler} (bc);
\end{tikzpicture}
\caption{Overview of the pipeline. The top row shows the artifacts derived from the Rocq source. The \encore{} compiler, run by \code{build.rs}, turns the optimized CPS code into bytecode that the embedded \encore{} VM executes inside the host firmware.}
\label{fig:pipeline}
\end{figure}

We first present an architecture for structuring firmware so its core is fully verifiable, illustrated on a concrete transaction-signing application.
Next, we introduce \encore{}, the runtime we built to execute this code on microcontrollers.
We then survey the existing extraction paths from theorem provers to embedded targets, and characterize why they fall short on this class of hardware.
Finally, we evaluate the pipeline on eight firmware workloads and outline remaining steps toward extending trust further up the toolchain.

This work makes three contributions:
\begin{itemize}
  \item We show that code extracted from Rocq can run on Cortex-M microcontrollers, by choosing Scheme as the extraction target and building a purpose-built bare-metal runtime.
  \item We present \encore, a \code{no\_std} CPS bytecode VM in Rust that fits within the resource budget of ST33-class hardware and executes unmodified Rocq-extracted Scheme.
  \item We introduce a firmware architecture structured as a pure state-transition function, making the business-logic core fully provable in Rocq while confining the unverified trusted computing base (TCB) to the event/effect boundary, which grows with the hardware interface rather than with application logic.
\end{itemize}

\section{An Architecture for Verified Firmware}
\label{sec:firmware}

This section answers \emph{what} to run on the MCU and how to structure it so correctness is fully provable in Rocq.

\subsection{The State Transition Model}

The model below is not imported from an existing specification: we define it for this work.
Its shape follows the well-established reactive state-machine pattern (an application reduced to a transition over an explicit state driven by input events), which we instantiate for firmware.
We choose this formulation because it maximizes the area that can be proven: modeling the application as a pure state transition function, effectively treating the whole embedded device as a state machine, confines all impurity to the event and effect boundary and leaves the transition itself as pure Gallina we can reason about directly.

Before diving into the modeling details, we assume that a Scheme VM is deployed as part of what we call a Formal Software Development Kit (FSDK) that bridges the virtual machine and the host device.
The FSDK contains both host code (it \emph{is} the real firmware that gets deployed) and Rocq code to provide the necessary bridge abstractions.

With this in mind, we define the state transition function as:
\[
  \code{step} : \code{State} \times \code{Event} \rightarrow \code{State} \times \code{list}\ \code{Effect}.
\]

Given the types:
\begin{description}
  \item[\textbf{State}] represents the application state. This will depend on what problem the application is solving, independently from the FSDK.
  \item[\textbf{Event}] is an incoming event from the device. Device events are linked to the underlying hardware and can be network packets, user interface signals such as pressed buttons or specific touch screen positions, etc.
  \item[\textbf{Effect}] is the description of an effectful computation that needs to be interpreted to be realized. A critical part of the FSDK is then to interpret the effects to perform them.
\end{description}

At run time, the FSDK loops: it polls an event $E_i$, computes $(S_{i+1}, F_i) = \code{step}(S_i, E_i)$, performs the effects $F_i$, and repeats with the new state.

This model applies two key functional programming principles, copying instead of mutating and describing effects rather than performing them directly, to firmware code, effectively maximizing the provable business logic part.

The remaining parts, the \textbf{Event poller} and the \textbf{Effect interpreter}, form the TCB.
Their size is independent of firmware complexity unless new hardware operations and capabilities are added.
The verified core grows with the application; the unverified boundary grows only with the hardware interface.

Moreover, in a VM-based deployment the application logic can be fully specified in the VM input code, allowing the same carefully reviewed base firmware to host custom, verified business logic, which is a particularly powerful combination for safety-critical devices.

\subsection{Admitted Axioms and the Trust Chain}
\label{sec:axioms}

Some operations are observationally pure from Rocq's perspective but would be impractical to implement fully from a business point of view.
Examples include hash functions, polynomial operations (CRC), and more generally any finite field operation involved in cryptographic algorithms.

Rather than relying on an Effect/Event pair with little added value, we choose to admit them in Rocq as axioms and provide them as host callbacks before the application run loop starts.
VM primitive operations, such as the byte-string operations \code{bytes\_len} and \code{bytes\_concat}, are treated the same way: Rocq parameters with axiomatized equations, mapped to the VM's implementations by extraction directives.
This bridge is manual: each host primitive is declared in Rocq as an axiom, with its behavior stated as a further axiom when a proof needs it, and linked to its Rust callback by an extraction directive, which keeps the trusted axioms few and auditable at the cost of keeping both sides in sync by hand.

The FSDK can also fix what every application must prove.
A typeclass over \code{State}, \code{Event}, and \code{step} can request a baseline set of theorems, for example that no signing effect is emitted outside an approved state; an application is then accepted only with an instance of that class, and reviewing a firmware application amounts to reviewing the statements of the class, which the Rocq kernel checks for every instance.

\subsection{Trust Assumptions}
\label{sec:trust}

The verified part of the system is the Rocq model of the application logic and its stated properties.
The trusted computing base includes the Rocq kernel and extraction mechanism, the Scheme frontend, the \encore{} compiler and VM, the Rust compiler, host callbacks implementing admitted axioms, and the event poller/effect interpreter supplied by the device firmware.
This boundary is deliberate: application growth extends the proved \code{step} function and its lemmas, whereas the unverified boundary changes only when new hardware actions, externs, or runtime primitives are added.
The current work therefore establishes verified firmware logic running on bare metal; Sect.~\ref{sec:semantics} states a compiler correctness conjecture whose proof, together with a proof that the Rust interpreter implements its abstract machine, would remove the compiler and VM from this list; both are future work, discussed in Sect.~\ref{sec:discussion}.

\subsection{Concrete Example: Transaction Signing Application}

As a concrete example, we built a Ledger Flex transaction-signing app with \encore.

LedgerOS provides us with higher-level primitives we used as our \textbf{Event poller} and our \textbf{Effect interpreter}.
We tailored the component we called FSDK to the particular use cases of this app for the sake of simplicity.
In a real implementation of a Formal SDK for this class of Ledger device and OS, the list of events and effects is expected to be much longer.

The device receives raw Application Protocol Data Unit (APDU) bytes over USB, parses them into structured fields, displays them for user approval, then signs and returns the result.
The app is a toy: it is not tied to any real crypto protocol.

\par\vspace{0.4em}\noindent
\begin{minipage}[t]{0.46\textwidth}
\begin{lstlisting}[language=Coq]
(* Application state *)
Inductive State : Type :=
  | InitialState
  | PendingTx (tx : Bytes).
(* Formal SDK (truncated) *)
Inductive Event : Type :=
  | InApdu (bytes : Bytes)
  | ApprovedTx
  | RejectedTx.
\end{lstlisting}
\end{minipage}\hfill
\begin{minipage}[t]{0.46\textwidth}
\begin{lstlisting}[language=Coq,firstnumber=last]
Inductive Effect : Type :=
  | OutApdu (bytes : Bytes)
  | DisplayProps
      (to : Bytes)
      (value : Bytes).
\end{lstlisting}
\end{minipage}
\par\vspace{0.4em}

Parser correctness is where formal verification pays off most directly.
Two properties are proven for our app:

\par\vspace{0.4em}\noindent
\begin{minipage}{\textwidth}
\begin{lstlisting}[language=Coq]
Theorem deserialize_preserves_data : forall data tx,
  deserialize_transaction data = inr tx ->
  tx_nonce tx ++ tx_to tx ++ tx_value tx ++ tx_memo tx
    = data.

Theorem check_encoding_spec : forall memo,
  check_encoding memo = true <->
  (forall b, In b memo -> b <= MAX_ASCII).
\end{lstlisting}
\end{minipage}
\par\vspace{0.4em}

\code{deserialize\_preserves\_data} guarantees that no bytes are dropped or reordered during parsing, a property a hardware wallet must ensure before signing.
\code{check\_encoding\_spec} gives a machine-checked certificate that memo content is valid ASCII.
Note that neither property is expressible as a Rust type.

These two functions are the parser and validator that \code{step} runs on an \code{InApdu} event: on success, \code{step} moves from \code{InitialState} to \code{PendingTx} only if \code{check\_encoding} holds, and emits a \code{DisplayProps} effect; every other input leaves the state unchanged.
The two theorems therefore constrain the behavior of the state-transition function on the event that admits untrusted input.
This is the point where the proofs meet the firmware.

\section{The \encore{} VM and Compiler}

\encore{} is a Scheme runtime built for the code Rocq extraction emits.
That code is heavily curried and close to CPS form, which suits a register-based virtual machine: the CPS rewrite eliminates the call stack \emph{semantically} and the register VM eliminates it \emph{physically}, so there is no hidden control flow and no frame management at run time, and every intermediate value is named, which makes register allocation direct.
Extracted code relies on a \code{macros\_extr.scm} file of macros for curried definition, application, and constructor matching; rather than support Scheme macros in general, \encore{} implements these as core primitives.
Its only addition to Scheme is the \code{(extern (slot $N$) \dots)} form, which declares a function provided by the host: Rocq poses it as an axiom, and the host registers it in slot $N$ before running the program.
We use a dedicated form because standard Scheme has no portable notion of a host-provided function, and the FFI forms of full implementations presuppose a hosted C runtime and dynamic loading.

\subsection{VM Internals}

\encore{} is a \code{\#![no\_std]} register-based bytecode interpreter with 256 registers and no call stack.
Every value is a packed 32-bit word encoding its type, metadata, and either an inline payload (integers, code pointers) or a heap address (closures, constructors, byte strings) (Fig.~\ref{fig:values}).

\begin{figure}[tbp]
\centering
\colorlet{vheap}{orange!18}
\colorlet{vcode}{blue!12}
\colorlet{vnum}{yellow!20}
\colorlet{vtyp}{gray!25}
\begin{tikzpicture}[x=0.29cm, y=0.36cm, font=\scriptsize]
  \newcommand\fld[5]{%
    \draw[fill=#4] (#2,-#1) rectangle ++(#3,-0.85);
    \node at ({#2+#3/2},{-#1-0.425}) {#5};}
  \newcommand\row[2]{\node[anchor=east] at (-0.4,{-#1-0.425}) {#2};}
  \newcommand\grp[2]{\node[anchor=west, font=\scriptsize\bfseries] at (-9,{-#1-0.45}) {#2};}
  \foreach \b in {31,24,16,8,0} {
    \node[font=\tiny, text=gray] at ({31-\b+0.5},0.45) {\b};
  }
  \grp{0}{Values (registers, globals, fields)}
  \row{1}{Integer}       \fld{1}{0}{24}{vnum}{signed integer, 24 bits}  \fld{1}{24}{8}{vtyp}{\texttt{04}}
  \row{2}{Function}      \fld{2}{0}{16}{vcode}{code address}            \fld{2}{16}{8}{white}{--}  \fld{2}{24}{8}{vtyp}{\texttt{05}}
  \row{3}{Closure}       \fld{3}{0}{16}{vheap}{heap address}            \fld{3}{16}{8}{white}{--}  \fld{3}{24}{8}{vtyp}{\texttt{00}}
  \row{4}{Constructor}   \fld{4}{0}{16}{vheap}{heap address or \texttt{NULL}} \fld{4}{16}{8}{white}{tag} \fld{4}{24}{8}{vtyp}{\texttt{01}}
  \row{5}{Bytes}         \fld{5}{0}{16}{vheap}{heap address}            \fld{5}{16}{8}{white}{--}  \fld{5}{24}{8}{vtyp}{\texttt{06}}
  \grp{6}{Headers (first words of a heap object)}
  \row{7}{GC header}     \fld{7}{0}{16}{vheap}{forwarding address}      \fld{7}{16}{8}{vnum}{mark, size}  \fld{7}{24}{8}{vtyp}{\texttt{03}}
  \row{8}{Closure header}\fld{8}{0}{16}{vcode}{code address}            \fld{8}{16}{8}{vnum}{env\_len}    \fld{8}{24}{8}{vtyp}{\texttt{02}}
  \row{9}{Bytes header}  \fld{9}{0}{24}{vnum}{byte length, 24 bits}     \fld{9}{24}{8}{vtyp}{\texttt{07}}
\end{tikzpicture}

\caption{\encore{} values: every value and heap header is one 32-bit word, \code{[payload:16 | meta:8 | typ:8]}. Integers, functions, and nullary constructors never allocate; heap addresses are 16-bit word indices. A heap object is a GC header followed by its contents: a closure header and the captured values, the fields of a constructor, or a bytes header and the bytes, 4 per word. A value points at its object's GC header, so \code{FIELD $i$} reads word $\mathit{addr} + 1 + i$.}
\label{fig:values}
\end{figure}

The 256 registers comprise \code{SELF} (current function), \code{CONT} (current continuation), the arguments \code{A1}--\code{A8} (\code{0x02}--\code{0x09}), general-purpose registers, and the \code{NULL} sentinel (\code{0xff}).
The heap is a bump-allocated arena with a mark-compact (Lisp-2) collector; since every value and header word carries its type byte (byte-string payloads are skipped by their header), the collector finds pointers without compiler-emitted maps.
The only call instruction is ENCORE, which sets \code{SELF} and \code{CONT} and jumps without returning; resuming a continuation is the same instruction with \code{NULL} as continuation, and MATCH and BRANCH provide all other control flow.
Sect.~\ref{sec:semantics} makes this precise.

\subsection{Compiler Pipeline}

The parser recognizes the extraction macros and produces a direct-style IR.
A first pass uncurries saturated applications, which are frequent in extracted code, saving the allocation of intermediate closures.
Variables are then replaced by de Bruijn indices, and the program is converted to CPS.
The CPS IR is optimized by iterating shrinking and growth-enabling passes to a fixed point, in the style of Appel~\cite{appel1992cwc}, before register allocation, a peephole pass, and bytecode emission.

\subsection{Host--VM Interface}

\encore{} is a library embedded in a Rust application that keeps control of the execution context, memory budget, and I/O.
The VM exposes two entry points: calling a global or a closure handle with typed arguments, and registering typed host callbacks invoked by EXTERN.
Values crossing the boundary are converted by the procedural macros \derive{ValueEncode} and \derive{ValueDecode}, whose \code{\#[ctor(tag)]} annotations use constructor tags from a \code{bindings.rs} file that \code{build.rs} generates together with the bytecode, so host code and bytecode cannot disagree on tags.
\code{VmList<T>} and \code{VmBytes} give bounded, allocation-free access to lists and byte strings in the VM heap.


\newcommand{\kw}[1]{\mathsf{#1}}
\newcommand{\instr}[1]{\textsc{#1}}

\subsection{Formal Semantics of the VM}
\label{sec:semantics}

We give the semantics of the bytecode machine, following the standard treatment of compilation to a virtual machine~\cite{leroy2009mechanized}.
It is transcribed from the interpreter and not yet mechanized.
$h_j$ is the host function registered in extern slot $j$, which may read and allocate on the heap, and $\delta_{op}$ the meaning of a primitive opcode, a partial function undefined exactly where the VM traps (integer overflow, out-of-range byte).

\paragraph{Bytecode machine.}
Fig.~\ref{fig:vm} gives the abstract machine that \code{encore\_vm} implements.
It abstracts the 32-bit packed encoding of Fig.~\ref{fig:values} into tagged words $w$ and the arena into a partial map $H$ from addresses to objects, and leaves the code $c$ implicit: $c(pc)$ is the instruction at $pc$ and $pc^{+}$ the next one.
A state is $\langle pc, R, H \rangle$, with $R$ the 256-register file. As in the implementation, the globals live in the same memory as the heap: global $i$ is a reserved cell $g_i$ of $H$ holding a word.
There is no stack component: $\instr{encore}$ is the only transfer to a code address held in a value, and no instruction returns.
When allocation finds no free address, the collector replaces $H$ by $\pi(H|_{\mathit{reach}})$ and $R$ by $\pi(R)$, where $\mathit{reach}$ is the set of addresses reachable from $R$ and the global cells, and $\pi$ is an injective renaming of these addresses that fixes the $g_i$; the machine traps with \code{HeapOverflow} if allocation still fails.
The collector is thus invisible up to renaming, which is the property a proof of the mark-compact implementation must establish.

\begin{figure}[!tbp]
\centering\small
\[
\resizebox{\ifdim\width>\linewidth\linewidth\else\width\fi}{!}{$\displaystyle
\begin{array}{r@{\;}c@{\;}l}
w & ::= & \kw{int}(n) \mid \kw{func}(\ell) \mid \kw{clos}(a) \mid \kw{con}(t, a) \mid \kw{con}(t, \bullet) \mid \kw{bytes}(a) \\
H(a) & ::= & \langle \ell; w_0, \dots, w_{m-1} \rangle \ \text{(closure)} \mid \langle w_0, \dots, w_{m-1} \rangle \ \text{(fields)} \mid s \ \text{(byte string)} \mid w \ \text{(global cell)}
\end{array}
\qquad
\begin{array}{l}
\mathit{code}_H(\kw{func}(\ell)) = \ell \\
\mathit{code}_H(\kw{clos}(a)) = \ell \ \text{if}\ H(a) = \langle \ell; \bar{w} \rangle
\end{array}
$}
\]
\vspace{-0.3em}
\[
\resizebox{\ifdim\width>\linewidth\linewidth\else\width\fi}{!}{$\displaystyle
\begin{array}{@{}l@{\quad}l@{\quad}l@{}}
\toprule
c(pc) & \text{condition} & \text{next state} \\
\midrule
\instr{mov}\ d\ r & & \langle pc^+, R[d \mapsto R(r)], H \rangle \\
\instr{capture}\ d\ i & R(\kw{SELF}) = \kw{clos}(a),\ H(a) = \langle \ell; \bar{w} \rangle & \langle pc^+, R[d \mapsto w_i], H \rangle \\
\instr{global}\ d\ i & H(g_i) = w & \langle pc^+, R[d \mapsto w], H \rangle \\
\instr{function}\ d\ \ell & & \langle pc^+, R[d \mapsto \kw{func}(\ell)], H \rangle \\
\instr{closure}\ d\ \ell\ \bar{r} & a \notin \mathrm{dom}(H) & \langle pc^+, R[d \mapsto \kw{clos}(a)], H[a \mapsto \langle \ell; R(\bar{r}) \rangle] \rangle \\
\instr{pack}\ d\ t\ \bar{r} & \text{arity}(t) > 0,\ a \notin \mathrm{dom}(H) & \langle pc^+, R[d \mapsto \kw{con}(t, a)], H[a \mapsto \langle R(\bar{r}) \rangle] \rangle \\
\instr{pack}\ d\ t & \text{arity}(t) = 0 & \langle pc^+, R[d \mapsto \kw{con}(t, \bullet)], H \rangle \\
\instr{field}\ d\ r\ i & R(r) = \kw{con}(t, a),\ H(a) = \langle \bar{w} \rangle & \langle pc^+, R[d \mapsto w_i], H \rangle \\
\instr{unpack}\ d\ t\ r & R(r) = \kw{con}(t', a),\ H(a) = \langle \bar{w} \rangle & \langle pc^+, R[d{+}i \mapsto w_i]_{i < \text{arity}(t)}, H \rangle \\
\instr{match}\ r\ b\ \ell_0 \dots \ell_{m-1} & R(r) = \kw{con}(t, \_),\ 0 \le t - b < m & \langle \ell_{t-b}, R, H \rangle \\
\instr{encore}\ f\ k & \mathit{code}_H(R(f)) = \ell \in \mathrm{dom}(c) & \langle \ell, R[\kw{SELF} \mapsto R(f), \kw{CONT} \mapsto R(k)], H \rangle \\
\instr{extern}\ d\ r\ j & h_j(R(r), H) = (w, H') & \langle pc^+, R[d \mapsto w], H' \rangle \\
\instr{op}\ d\ \bar{r} & \delta_{op}(R(\bar{r}), H) = (w, H') & \langle pc^+, R[d \mapsto w], H' \rangle \\
\instr{fin}\ r & & \text{halt with } R(r) \\
\bottomrule
\end{array}
$}
\]
\caption{Bytecode machine (\code{vm.rs}). $\instr{branch}\ r\ b\ \ell_0\ \ell_1$ jumps to $\ell_0$ if the tag of $R(r)$ is $b$ and to $\ell_1$ otherwise; $\instr{global\_w}$ is $\instr{global}$ with a 16-bit index; $\instr{op}$ stands for the literal, integer, and byte-string opcodes, where $H'$ extends $H$ when the result is a new byte string. A state with no applicable rule is stuck. The interpreter traps on some stuck states (unknown tag, call outside the code, arithmetic overflow) but does not check value types: compiled code must never reach them.}
\label{fig:vm}
\end{figure}

\paragraph{Entry points.}
Code address $0$ holds the return stub $\instr{fin}\ \kw{A1}$.
The host calls a function value $f$ on arguments $\bar{w}$ by running the machine from $\langle \mathit{code}_H(f), R_0[\kw{SELF} \mapsto f, \kw{CONT} \mapsto \kw{func}(0), \kw{A}i \mapsto w_i], H \rangle$, with $R_0$ the cleared register file; the call returns $w$ when the machine halts with $w$.
Loading calls the entry point of each definition in order and stores the result in its global cell.

\paragraph{Correctness statement.}
Let $v \approx_H w$ relate first-order values to machine words: $n \approx_H \kw{int}(n)$; $s \approx_H \kw{bytes}(a)$ when $H(a) = s$; $C() \approx_H \kw{con}(t(C), \bullet)$; and $C(\bar{v}) \approx_H \kw{con}(t(C), a)$ when $H(a) = \langle \bar{w} \rangle$ and $v_i \approx_H w_i$ for all $i$, where $t(C)$ is the tag of constructor $C$.

\begin{conjecture}[Compiler correctness]
\label{conj:correct}
Let $P$ be an extracted program accepted by the frontend and $B$ the bytecode \encore{} produces for it.
If a definition $x$ of $P$ evaluates, under the call-by-value semantics of Scheme, to a first-order value $v$ (built from integers, byte strings, and constructors), then loading $B$ either initializes the global cell $g_x$ of $x$ to a word $w$ such that $v \approx_H w$ in the resulting heap $H$, or traps with \code{HeapOverflow}, \code{IntOverflow}, or \code{ByteRange}.
Likewise, if $x$ denotes a function and $x\ v_1 \dots v_m$ evaluates to a first-order $v$ for first-order arguments $v_i \approx_H w_i$, then calling $x$ on $\bar{w}$ returns a word $w$ with $v \approx_{H'} w$ in the final heap $H'$, or traps likewise.
\end{conjecture}

\noindent
Since the machine is deterministic, this forward simulation also excludes any other observable behavior~\cite{leroy2009mechanized}.
The statement decomposes along the pipeline into three simulations, each with a precedent in a mechanized compiler:
(i)~the frontend and the CPS transformation, as CertiRocq does for its front-end passes such as closure conversion~\cite{paraskevopoulou2019closure};
(ii)~each optimization pass, as in Pilsner~\cite{neis2015pilsner};
(iii)~register allocation and emission against the machine of Fig.~\ref{fig:vm}, as CertiRocq does for C~\cite{savarybelanger2019cpstoc}.
Two links remain outside the statement: Rocq's extraction, for which a verified alternative exists for OCaml~\cite{forster2024verified}, and the agreement between Fig.~\ref{fig:vm} and the Rust interpreter, discussed in Sect.~\ref{sec:discussion}.

\FloatBarrier

\section{Related Work}
\label{sec:related}

Our targets are ST33 secure microcontrollers (Cortex-M35P): the ST33K1M5 has 64\,KiB of RAM and 1.5\,MiB of flash, and the tighter ST33J2M0, 50\,KiB of RAM, which sets the binding constraint.
Applications on this class of device ship with no runtime and cannot rely on libc, libm, or the Rust standard library, which excludes many proof-to-code paths.

Lean 4~\cite{demoura2021lean4} has mature native code generation, but its runtime assumes C++ library support, reference counting, tasks, and hosted I/O, and no port targets bare-metal devices of this size.
Rocq extraction to OCaml~\cite{letouzey2008extraction} is mature, but the OCaml runtime requires a generational GC, libc/libm support, and native-code assumptions mismatched with Cortex-M object formats; OMicroB~\cite{OMicroB}, a bytecode VM for microcontrollers, supports AVR and PIC32 but not the full Thumb-2 of our targets.

CertiRocq (formerly CertiCoq)~\cite{anand2017certicoq} is the most principled Rocq compilation path: it compiles Gallina to Clight through a largely verified compiler chain.
Its runtime does not carry over unchanged to our targets.
The generated Clight allocates inductive values, and correctness is stated against the heap of its own collector; on a microcontroller, the collector must run over a static arena, outside that proof.
CertiRocq also ignores Rocq's extraction directives: mapping \code{nat} to machine integers requires registering primitives, applied by an erasure pass that is itself trusted, and its code is in direct style, so deep recursion consumes the C stack (Sect.~\ref{sec:eval}).

Rocq also supports extraction to Scheme \cite{letouzey2008extraction}.
Although less mature than OCaml extraction, Scheme has compact semantics and no mandatory hosted runtime.
Small Scheme systems such as Ribbit~\cite{yvon2021ribbit} show that complete implementations fit microcontroller budgets, but none accepts the code Rocq extracts: Ribbit, for instance, does not support the quasiquote that extraction emits to build constructors.
\encore{} is a runtime built for that code.

Verified compilers such as CompCert \cite{leroy2009compcert} and CakeML \cite{kumar2014cakeml} carry a machine-checked proof that the generated code refines its source, and \OE{}uf \cite{mullen2018oeuf} applies this idea to a subset of Gallina to shrink the extraction TCB.
Another line of work obtains verified low-level code by restricting how programs are written: Low*~\cite{protzenko2017lowstar} generates C without garbage collection from a first-order subset, and Bedrock2 \cite{erbsen2021bedrock2} verifies a low-level language end-to-end on a small RISC-V embedded system.
These approaches give stronger guarantees than ours, but either constrain the source to first-order code without garbage collection or assume a hosted runtime.
\encore{} is not a new verification methodology: it is a practical systems pipeline that runs ordinary higher-order, allocating Gallina, produced by the standard extraction mechanism, on Cortex-M hardware, with the remaining compiler/VM gap made explicit in Sect.~\ref{sec:trust}.

Table~\ref{tab:comparison} summarizes how these paths meet our constraints.
Lean 4 and OCaml are eliminated by their runtimes.
Existing small Scheme systems do not accept extracted code.
This leaves \encore{} and CertiRocq, which we compare in Sect.~\ref{sec:eval}.
Except for CertiRocq, proofs cover only the source program: \encore{}'s advantage is deployability, not a stronger guarantee (Sect.~\ref{sec:trust}).

\begin{table}[t]
\centering
\caption{Rocq extraction paths against embedded deployment constraints.}
\label{tab:comparison}
\resizebox{\textwidth}{!}{%
\begin{tabular}{lllll}
\toprule
Extraction path & GC required & \code{no\_std} viable & Cortex-M viable & Verified down to \\
\midrule
Lean 4 & Reference counting & No & No & Source \\
OCaml / OMicroB & Generational GC & No & No & Source \\
CertiRocq & Generational, static arena & Patched runtime & Small inputs & Clight \\
Scheme / \encore{} & Mark-compact, bounded & Yes & Yes & Source \\
\bottomrule
\end{tabular}
}
\end{table}

\section{Evaluation}
\label{sec:eval}

We evaluate the pipeline against four research questions.
\textbf{RQ1 (memory):} which workloads fit a budget of 50\,KiB of RAM (the ST33J2M0's) and 256\,KiB of flash, for each toolchain?
\textbf{RQ2 (cost):} what does \encore{} cost in executed instructions and flash, against native Rust and against CertiRocq?
\textbf{RQ3 (predictability):} is the latency added by the garbage collector bounded?
\textbf{RQ4 (proof effort):} what is proved, with how much proof, and how much unverified code remains?
The workloads, drivers, measurement harness, and raw results are public~\cite{encorebench}.

\subsection{Methodology}

\textbf{Variants.}
\textbf{E} is the pipeline of this paper (\encore{} 0.1.6, CPS optimizer on).
\textbf{C} compiles the same Gallina with CertiRocq 0.9.1 to Clight and \code{gcc -Os}, with its generational collector running over a static arena and \code{nat} mapped to machine integers, as for \encore{}.
\textbf{R} is idiomatic \code{no\_std} Rust without an allocator, written without looking at the Gallina: it is the performance ceiling and the output oracle, as every E and C run must reproduce R's output hash.
The three variants share the Rust driver, the frozen input vectors, and the output hashing; only the logic differs.

\textbf{Workloads.}
Table~\ref{tab:workloads} describes the eight workloads and the property proved for each.
We chose them because a bug in them is a security hole or a field failure on a hardware wallet, and because a precise property can be proved about them.
Each follows the \code{step} model of Sect.~\ref{sec:firmware} and is parameterized by an input size $N$.
W8 is deliberately unfavorable to \encore{}: a tight bitwise loop that allocates almost nothing, leaving no allocation to amortize interpretation.

\textbf{Platform and metrics.}
Linker scripts impose the ST33J2M0 budget, so a program that does not fit fails to link.
Within it, E uses a 32\,KiB heap (40\,KiB for W6) and C a 20\,KiB arena, the rest of the RAM holding its C stack.
We run on QEMU 8.2.2, emulating a Cortex-M33 (AN505), the ARMv8-M core closest to the Cortex-M35P; instruction counts on a Cortex-M3 (LM3S6965) differ by at most 1.3\%.
The metric for speed is the number of instructions executed per run, counted exactly by tracing QEMU between two markers, minus the cost of an empty region.
We also report the flash size of the linked image, the smallest E heap that completes every case (by binary search), and, in a separate build since its counters perturb the instruction counts, the number, share, and longest pause of collections.
The other tools are Rust 1.88, \code{arm-none-eabi-gcc} 13.2.1, and Rocq 9.1.1.

\begin{table}[t]
\centering
\caption{Benchmark description. Proof: number of theorems and lines of proof. Prog.: lines of program and specification (non-blank, comments excluded). No lemma is admitted. $N$: range of input sizes (bytes, events, rules, or entries).}
\label{tab:workloads}
{\scriptsize
\setlength{\tabcolsep}{4pt}
\begin{tabular}{@{}l>{\raggedright\arraybackslash}p{4.6cm}rrrr@{}}
\toprule
 & & \multicolumn{2}{c}{Proof} & & \\
\cmidrule(lr){3-4}
Workload & Property proved & thms & lines & Prog. & $N$ \\
\midrule
W1 APDU/TLV & totality, round trip & 25 & 240 & 217 & 5--261 \\
W2 RLP decoder & displayed fields are the signed ones & 24 & 284 & 243 & 0--260 \\
W3 BIP32 policy & signed iff a rule allows it & 30 & 216 & 234 & 1--64 \\
W4 PIN machine & counter never rises without a verify & 6 & 35 & 106 & 1--1000 \\
W5 A/B update & no rollback; a valid slot always exists & 14 & 71 & 218 & 10--1000 \\
W6 COBS & round trip; no zero byte in output & 12 & 101 & 105 & 16--1024 \\
W7 cred.\ store & balanced tree; lookup after insert & 27 & 163 & 150 & 10--500 \\
W8 CRC & equals the polynomial definition & 48 & 313 & 178 & 16--1024 \\
\bottomrule
\end{tabular}}
\end{table}

\subsection{Results}

This section reports our measurements, organized by research question.
Every E and C run reproduced R's output hash, so all variants compute the same results.
Table~\ref{tab:results} reports, for each workload, the memory results (RQ1) and the executed instructions and flash size (RQ2).
RQ3 relies on the separate collector build, and RQ4 on the proof figures of Table~\ref{tab:workloads}.

\begin{table}[!h]
\centering
\caption{Results. \emph{Largest $N$}: largest input that completes within the ST33J2M0 budget; E completes every tested size. \emph{Heap}: smallest E heap that completes every size (C uses a fixed 20\,KiB arena). Instructions and flash are measured at C's largest $N$ (W8: E's). \emph{R}: instructions executed by native Rust; E/R, C/R, E/C: ratios of executed instructions. Flash of the linked image; E's bytecode in parentheses. --: C not built for W8; W4 heap not measured.}
\label{tab:results}
{\scriptsize
\setlength{\tabcolsep}{3.5pt}
\begin{tabular}{@{}lrrrrrrrrrr@{}}
\toprule
 & \multicolumn{2}{c}{Largest $N$} & Heap & \multicolumn{4}{c}{Instructions} & \multicolumn{3}{c}{Flash (KiB)} \\
\cmidrule(lr){2-3}\cmidrule(lr){4-4}\cmidrule(lr){5-8}\cmidrule(l){9-11}
Workload & E & C & (KiB) & R & E/R & C/R & E/C & R & C & E (bytecode) \\
\midrule
W1 APDU/TLV     & 261  & 32  & 10.5  & 331      & 448       & 68 & 6.6  & 9.7  & 17.9 & 36.1 (3.7) \\
W2 RLP decoder  & 260  & 68  & 16.75 & 3{,}659  & 191       & 32 & 6.1  & 10.7 & 24.1 & 37.7 (5.1) \\
W3 BIP32 policy & 64   & 16  & 1     & 15{,}335 & 589       & 59 & 10.0 & 15.5 & 21.0 & 40.9 (3.1) \\
W4 PIN machine  & 1000 & 100 & --    & 7{,}003  & 208       & 24 & 8.5  & 15.8 & 22.5 & 40.9 (2.7) \\
W5 A/B update   & 1000 & 100 & 28    & 17{,}152 & 155       & 16 & 10.0 & 20.8 & 31.7 & 49.2 (6.5) \\
W6 COBS         & 1024 & 64  & 32.25 & 1{,}490  & 187       & 39 & 4.8  & 10.4 & 16.1 & 35.5 (2.0) \\
W7 cred.\ store & 500  & 50  & 12.75 & 22{,}379 & 34        & 4  & 8.1  & 12.9 & 16.3 & 37.4 (3.4) \\
W8 CRC          & 1024 & --  & 0.5   & 15{,}392 & 1{,}849   & -- & --   & 11.4 & --   & 34.9 (1.9) \\
\bottomrule
\end{tabular}}
\end{table}

\textbf{RQ1: memory.}
E completes all 31 cases within the budget, with minimum heaps from 0.5 to 32.25\,KiB.
C completes 17 of the 27 cases of W1--W7.
Its generated code is in direct style and recurses on the C stack once per input element or event.
The larger inputs therefore exhaust the arena or the stack.
CertiRocq's CPS mode would move this recursion to the heap, as in \encore{}.
In version 0.9.1, however, it cannot be combined with the mapping of \code{nat} to machine integers.

\textbf{RQ2: cost.}
Interpretation is expensive.
At C's largest $N$, E executes 4.8--10$\times$ more instructions than C (2.3--10$\times$ over all inputs).
E executes 155--589$\times$ more instructions than R, with two exceptions: W7 (34$\times$), where tree manipulation also dominates R, and W8 (1{,}849$\times$), by design.
C itself executes 4--68$\times$ more instructions than R.
The CPS optimizer divides E's instruction count by 2.3--4.2 and halves its bytecode.
All images are far below the 256\,KiB flash budget.
E adds a constant 21--23\,KiB over R for the VM and its dependencies.
Its bytecode (1.9--6.5\,KiB) is two to three times smaller than R's native code for the policy workloads W3--W5.
The VM thus pays for itself in flash only for programs several times larger than ours.

\textbf{RQ3: predictability.}
The collector runs only when the heap is full, so a pause is bounded by the heap size, not by the input.
At the largest $N$, collections take 3--22\% of the execution, and the longest pause is 49--309 thousand instructions.
The counts are deterministic.
Turning them into a latency bound in time requires cycle measurements on hardware, which we have not yet done.

\textbf{RQ4: proof effort.}
Every workload proves its property in 35--313 lines of proof, for 105--243 lines of program and specification, with no admitted lemma (Table~\ref{tab:workloads}).
The unverified Rust that drives E stays at 93--104 lines across workloads.
In contrast, R's 68--238 lines (logic and driver) carry no proof at all.

\textbf{Hardware validation.}
The workloads above run on QEMU.
The transaction-signing application of Sect.~\ref{sec:firmware} was also validated on physical Ledger Flex hardware: the Rocq-extracted, \encore-compiled firmware parses, displays, and signs data on the device.
Its verified Gallina core (\code{step}, parser, helpers, and proofs) has about 400 lines, and its unverified Rust layer about 575 lines.

\section{Discussion}
\label{sec:discussion}

\subsection{Threats to Validity}

Instruction counts are not cycles.
The interpreter and native code differ in memory traffic, so the ratios of Table~\ref{tab:results} bound the gap only roughly.
The cycle-counter path for physical boards is implemented but not yet validated.
R was written by the authors.
C's arena is smaller than E's heap because the C stack shares the RAM budget.
Line counts are ours and approximate.

\subsection{Trace Properties}

The properties of Sect.~\ref{sec:firmware} constrain a single call to \code{step}.
Firmware also needs properties over event sequences: ordering constraints, authorization before a sensitive operation, or the absence of forbidden transitions.
The model expresses them by folding \code{step} over a list of events, which yields the reachable states and the emitted effects.
Safety properties then become invariants proved by induction on the event list, such as ``no signing effect without a prior \code{ApprovedTx}''.
The FSDK typeclass of Sect.~\ref{sec:axioms} can require such invariants from every application.
Liveness is out of scope: the device cannot force the host or the user to send an event.

\subsection{Verified Optimization Passes}

The nine CPS optimization passes are trusted Rust code.
Proving them semantics-preserving, step~(ii) of the decomposition of Conjecture~\ref{conj:correct}, is standard for the shrinking reductions, and the growth-enabling passes (inlining, hoisting, CSE, contification) resemble optimizations verified in Pilsner~\cite{neis2015pilsner}.

\subsection{Ahead-of-Time Native Compilation}

A backend compiling the CPS IR directly to ARMv8-M Thumb-2 is designed but not yet implemented: it pins the most used virtual registers to hardware registers and turns each tail call into a code-table lookup and a \code{BX}, removing interpreter dispatch.

\subsection{Portability}
\label{sec:portability}

Only the TCB boundary is device-specific: the \code{step} function, the bytecode, and \code{encore\_vm} (\code{no\_std}, no libc) are architecture-independent, so porting reduces to retargeting the Rust build and reimplementing the event poller, effect interpreter, and host callbacks (Sect.~\ref{sec:axioms}).
Any Cortex-M with a Rust target and a few tens of KiB of RAM qualifies, since the runtime adds about 22\,KiB of flash and our workloads need at most 32.25\,KiB of heap; this includes STM32 parts and ARM-based Arduino boards, but not 8-bit AVR boards, with 2\,KiB of RAM and only a tier-3 Rust target.

\subsection{Verifying the VM with rocq-of-rust}

\code{encore\_vm} is \code{no\_std} Rust with minimal \code{unsafe} and no trait objects in the interpreter loop.
The \code{rocq-of-rust} tool \cite{ho2024rocqofrust} translates Rust source to Rocq definitions, enabling correctness proofs of Rust programs inside Rocq.
Applying it to \code{encore\_vm} would close the VM's gap in the trust chain: the same proof assistant that certifies the firmware logic certifies the runtime executing it.

The central theorem to establish is the agreement of the Rust dispatch loop with the machine of Fig.~\ref{fig:vm}, the link left outside Conjecture~\ref{conj:correct}: each step of the interpreter corresponds to a transition of the abstract machine.
Two components present the main proof obligations.
The mark-compact GC requires reasoning about pointer aliasing and preservation of the tag-byte invariant across every collection cycle, since compaction rewrites every live pointer in both the register file and the heap interior.
The ENCORE dispatch opcode must be shown to resolve the callee's code pointer and to preserve the register-file state prescribed by the calling convention.
The \code{unsafe} surface is small: unchecked reads of code, whose bounds load-time validation establishes, of registers, in bounds by construction, and of the arena, which relies on the compiler emitting well-typed accesses.

\section{Conclusion}

We have shown that Rocq-proven firmware logic runs on a Cortex-M35P: extracted to Scheme, compiled by \encore, and deployed on a Ledger Flex hardware wallet.
The verified core remains the source of truth, while the runtime, compiler, host callbacks, and device I/O are explicit trusted components.
On eight security-relevant workloads, \encore{} fits the memory budget at every input size, whereas CertiRocq, on the seven where it was built, fails on the larger inputs; \encore{} pays for it in speed, with up to 10$\times$ more instructions than CertiRocq's native code.

The architecture is also designed to scale.
Structuring firmware as a pure state-transition function decouples verification effort from hardware complexity: the verified \code{step} grows with application logic, while the unverified components grow only with the hardware interface.

Proven application logic runs on bare metal today; closing the trust chain down to the VM and scaling to larger applications remain future work.

\section*{Acknowledgments}
The authors thank Nicolas Iooss for their valuable contribution and the Ledger Backend and Ledger Donjon teams for their support.

\bibliographystyle{splncs04}
\bibliography{refs}

@misc{encore,
  author = {Valentin Bergeron},
  title  = {{Encore!}: A Bytecode {VM} for {Rocq}-Extracted Code on Microcontrollers},
  year   = {2026},
  url    = {https://github.com/vbergeron/encore},
}

@book{appel1992cwc,
  author    = {Andrew W. Appel},
  title     = {Compiling with Continuations},
  publisher = {Cambridge University Press},
  year      = {1992},
  isbn      = {978-0-521-41695-5},
}

@misc{leroy2009mechanized,
  author       = {Xavier Leroy},
  title        = {Mechanized Semantics, with Applications to Program Proof and
                  Compiler Verification},
  howpublished = {Lecture notes, Marktoberdorf Summer School},
  year         = {2009},
  url          = {https://xavierleroy.org/courses/Marktoberdorf-2009/},
}

@article{paraskevopoulou2019closure,
  author    = {Zoe Paraskevopoulou and Andrew W. Appel},
  title     = {Closure Conversion Is Safe for Space},
  journal   = {Proc. ACM Program. Lang.},
  volume    = {3},
  number    = {ICFP},
  pages     = {83:1--83:29},
  year      = {2019},
  doi       = {10.1145/3341687},
}

@inproceedings{neis2015pilsner,
  author    = {Georg Neis and Chung-Kil Hur and Jan-Oliver Kaiser and
               Craig McLaughlin and Derek Dreyer and Viktor Vafeiadis},
  title     = {Pilsner: A Compositionally Verified Compiler for a
               Higher-Order Imperative Language},
  booktitle = {Proceedings of the 20th {ACM SIGPLAN} International Conference
               on Functional Programming ({ICFP} 2015)},
  pages     = {166--178},
  publisher = {ACM},
  year      = {2015},
  doi       = {10.1145/2784731.2784764},
}

@misc{savarybelanger2019cpstoc,
  author = {Olivier {Savary B{\'e}langer} and Matthew Z. Weaver and Andrew W. Appel},
  title  = {Certified Code Generation from {CPS} to {C}},
  year   = {2019},
  howpublished = {Manuscript},
  url    = {https://www.cs.princeton.edu/~appel/papers/CPStoC.pdf},
}

@article{forster2024verified,
  author    = {Yannick Forster and Matthieu Sozeau and Nicolas Tabareau},
  title     = {Verified Extraction from {Coq} to {OCaml}},
  journal   = {Proc. ACM Program. Lang.},
  volume    = {8},
  number    = {PLDI},
  year      = {2024},
}

@inproceedings{demoura2021lean4,
  author    = {Leonardo {de Moura} and Sebastian Ullrich},
  title     = {The {Lean} 4 Theorem Prover and Programming Language},
  booktitle = {Automated Deduction -- {CADE} 28},
  series    = {Lecture Notes in Computer Science},
  volume    = {12699},
  pages     = {625--635},
  publisher = {Springer},
  year      = {2021},
  doi       = {10.1007/978-3-030-79876-5_37},
}

@inproceedings{letouzey2008extraction,
  author    = {Letouzey, Pierre},
  title     = {Extraction in {Coq}: An Overview},
  booktitle = {Logic and Theory of Algorithms (CiE 2008)},
  year      = {2008},
  publisher = {Springer},
}

@misc{OMicroB,
  author = {Varoumas, Simon and Vaugon, Beno\^{i}t and Chailloux, Emmanuel},
  title  = {{OMicroB}: An {OCaml} Generic Virtual Machine for Microcontrollers},
  url    = {https://github.com/stevenvar/OMicroB},
  note   = {Accessed: May 2026},
}

@inproceedings{anand2017certicoq,
  author    = {Abhishek Anand and Andrew W. Appel and Greg Morrisett and
               Zoe Paraskevopoulou and Randy Pollack and
               Olivier {Savary B{\'e}langer} and Matthieu Sozeau and Matthew Weaver},
  title     = {{CertiCoq}: A Verified Compiler for {Coq}},
  booktitle = {Proceedings of the Workshop on Coq for Programming Languages ({CoqPL})},
  year      = {2017},
  note      = {Co-located with POPL 2017},
  url       = {https://www.cs.princeton.edu/~appel/papers/certicoq-coqpl.pdf},
}

@inproceedings{yvon2021ribbit,
  author    = {Samuel Yvon and Marc Feeley},
  title     = {A Small {Scheme} {VM}, Compiler, and {REPL} in {4K}},
  booktitle = {Proceedings of the 13th {ACM} {SIGPLAN} International Workshop on
               Virtual Machines and Intermediate Languages ({VMIL} 2021)},
  pages     = {14--24},
  publisher = {ACM},
  year      = {2021},
  doi       = {10.1145/3486606.3486783},
}

@article{leroy2009compcert,
  author    = {Xavier Leroy},
  title     = {Formal Verification of a Realistic Compiler},
  journal   = {Communications of the {ACM}},
  volume    = {52},
  number    = {7},
  pages     = {107--115},
  year      = {2009},
  doi       = {10.1145/1538788.1538814},
}

@inproceedings{kumar2014cakeml,
  author    = {Ramana Kumar and Magnus O. Myreen and Michael Norrish and Scott Owens},
  title     = {{CakeML}: A Verified Implementation of {ML}},
  booktitle = {Proceedings of the 41st {ACM} Symposium on Principles of
               Programming Languages ({POPL} 2014)},
  pages     = {179--191},
  publisher = {ACM},
  year      = {2014},
  doi       = {10.1145/2535838.2535841},
}

@inproceedings{mullen2018oeuf,
  author    = {Eric Mullen and Stuart Pernsteiner and James R. Wilcox and
               Zachary Tatlock and Dan Grossman},
  title     = {{\OE}uf: Minimizing the {Coq} Extraction {TCB}},
  booktitle = {Proceedings of the 7th {ACM SIGPLAN} International Conference
               on Certified Programs and Proofs ({CPP} 2018)},
  pages     = {172--185},
  publisher = {ACM},
  year      = {2018},
  doi       = {10.1145/3167089},
}

@article{protzenko2017lowstar,
  author    = {Jonathan Protzenko and Jean-Karim Zinzindohou{\'e} and
               Aseem Rastogi and Tahina Ramananandro and Peng Wang and
               Santiago Zanella-B{\'e}guelin and Antoine Delignat-Lavaud and
               C{\u{a}}t{\u{a}}lin Hri{\c{t}}cu and Karthikeyan Bhargavan and
               C{\'e}dric Fournet and Nikhil Swamy},
  title     = {Verified Low-Level Programming Embedded in {F*}},
  journal   = {Proc. ACM Program. Lang.},
  volume    = {1},
  number    = {ICFP},
  pages     = {17:1--17:29},
  year      = {2017},
  doi       = {10.1145/3110261},
}

@inproceedings{erbsen2021bedrock2,
  author    = {Andres Erbsen and Samuel Gruetter and Joonwon Choi and
               Clark Wood and Adam Chlipala},
  title     = {Integration Verification across Software and Hardware for a
               Simple Embedded System},
  booktitle = {Proceedings of the 42nd {ACM SIGPLAN} International Conference
               on Programming Language Design and Implementation ({PLDI} 2021)},
  pages     = {604--619},
  publisher = {ACM},
  year      = {2021},
  doi       = {10.1145/3453483.3454065},
}

@misc{encorebench,
  author = {Valentin Bergeron},
  title  = {{Encore!} Benchmarks: Workloads, Harness, and Results},
  year   = {2026},
  url    = {https://github.com/vbergeron/encore-benchmarks},
}

@misc{ho2024rocqofrust,
  author = {{Formal Land}},
  title  = {rocq-of-rust: Formal Verification of {Rust} Programs in {Rocq}},
  year   = {2024},
  url    = {https://github.com/formal-land/rocq-of-rust},
}

\end{document}